\documentstyle[12pt,aasms]{article}
\tightenlines

\def\gsim{{{}_>\atop{}^{{}^\sim}}}
\def\lsim{{{}_<\atop{}^{{}^\sim}}}
\def\chisq{\chi^{2}}
\def\h2{H$_{2}$}

\begin{document}

\begin{center}
Accepted for publicaton in {\it The Astrophysical Journal}
\end{center}

\vskip 24pt

\title{Near-Infrared Spectroscopy of Molecular \\
Filaments in the Reflection Nebula NGC 7023}

\vskip 24pt

\author{Paul Martini, K. Sellgren}

\vskip 12pt

\affil{Department of Astronomy, 174 W. 18th Ave., Ohio State University, \\
Columbus, OH 43210 \\
martini,sellgren@payne.mps.ohio-state.edu}

\and

\author{Joseph L. Hora}

\vskip 12pt

\affil{Institute for Astronomy, University of Hawaii, Honolulu, HI 96822 \\
hora@galileo.ifa.hawaii.edu}

\clearpage

\centerline
{\bf Abstract}

We present near-infrared spectroscopy of fluorescent molecular hydrogen (\h2)
emission from molecular filaments in the
reflection nebula NGC 7023. We derive the relative column densities of \h2
rotational-vibrational states from the measured line emission
and compare these results with several model photodissociation regions
covering a range of densities, incident UV-fields, and excitation
mechanisms. Our best-fit models for one filament suggest, but do not
require, either a combination of different densities, suggesting clumps of
10$^6$ cm$^{-3}$ in a $10^4 - 10^5$ cm$^{-3}$ filament, or a combination
of fluorescent excitation and thermally-excited gas, perhaps due to a
shock from a bipolar outflow. We derive densities and UV fields for
these molecular filaments that are in agreement with previous
determinations. 


\clearpage

\section{Introduction}

Near-infrared spectroscopic measurements of the relative line strengths
of molecular hydrogen (\h2) are a powerful way to determine the densities,
incident ultraviolet (UV) radiation field, and excitation mechanisms present in
photodissociation regions (PDRs).
\h2 emission, attributed to shocks, was first detected
in the planetary nebula NGC 7027 and the Orion Nebula (Treffers et al.  1976;
Gautier et al. 1976).
UV-pumped fluorescent molecular hydrogen was first detected in the
reflection nebula NGC 2023 (Gatley et al. 1987; Sellgren 1986).

Measurements of the relative and absolute intensities of
rotational-vibrational transitions in \h2 can be readily analyzed
with respect to existing models for collisional and fluorescent
excitation. These static models include UV-illuminated, low-density clouds
(Black \& van Dishoeck 1987), UV-illuminated, high-density clouds
(Sternberg \& Dalgarno 1989; Draine \& Bertoldi 1996), UV-illuminated, clumpy
clouds (Burton, Hollenbach, \& Tielens 1990), and shocked clouds (Hollenbach 
\& McKee 1979, 1989; Brand et al. 1988; Brand et al. 1989).
Application of PDR models to NGC 2023 confirmed that UV-pumped fluorescence
is likely the dominant excitation mechanism (Black \& van Dishoeck 1987;
Burton et al. 1990; Draine \& Bertoldi 1996).
These PDR models are applicable to a wide variety of objects such as
HII regions (e.g. Orion), planetary nebulae (e.g. NGC 7027) as well as
other reflection nebulae (Black \& van Dishoeck 1987; Sternberg \&
Dalgarno 1989).  Recent work
by Goldschmidt \& Sternberg (1995) has shown that time-dependent
PDR models may be required as well, a scenario these authors apply to
NGC 2023.

We present near-infrared spectroscopy of \h2 emission in NGC 7023, a
reflection nebula which is similar in many ways to NGC 2023.
One difference, however, is that NGC 2023 is illuminated by a
main-sequence B star, while NGC 7023 is illuminated by a
pre-main-sequence B star, the Herbig Ae/Be star HD 200775.

Chokshi et al. (1988) observed emission in atomic fine-structure lines
of NGC 7023, which represents the bulk of the cooling mechanism of most
PDRs (Tielens \& Hollenbach 1986; Hollenbach et al. 1991).
Their grid of measurements of [O I] (63 $\mu$m) and [C II] (158 $\mu$m), with
beam diameters of 33$''$ and 55$''$, respectively, were centered on a
region $30''$ N, $20''$ W of the central star. Their PDR model yields a
peak density of $\sim 4 \times 10^{3} $cm$^{-3}$ surrounded by a lower
density, $\sim 500$ cm$^{-3}$ region where CO (Watt et al.
1986) is formed. Based upon their line measurements, they also derive
an incident UV field of $G_{0} = 2.6 \times 10^3$ in the units of
Habing (1968) where Habing estimated the interstellar UV flux to be
$1.2 \times 10^{-4}$ ergs cm$^{-2}$ s$^{-1}$ sr$^{-1}$.

Rogers et al. (1995) compare H I, CO, and
IRAS observations of NGC 7023 taken with a resolution of approximately
1$'$. The peak in atomic hydrogen emission agrees
with the peak in the fine-structure lines observed by Chokshi et al.
(1988). However, Rogers et al. (1995) also find a molecular rim surrounding 
the atomic region and a high contrast in densities between the atomic and 
molecular regions. They find similar ($\sim
10^4$ cm$^{-3}$) molecular densities to those reported by
Chokshi et al. (1988) when using equilibrium PDR models. They also
compared their observations to time-dependent models and they find no evidence
for either a non-equilibrium PDR or an active bipolar outflow. Rather they 
propose that an outflow in the past paved the way for the development of the
current low-density PDR.

The presence of a bipolar outflow in NGC 7023 has long been a matter of
some debate. Watt et al. (1986) mapped the nebula in both the $^{12}$CO
and $^{13}$CO $J=1-0$ transitions with a half-power beamwidth of 1$'$
and found morphological features characteristic of an outflow source,
though they did
not have sufficient signal-to-noise to resolve high-velocity wings in
their spectra. They also report a molecular hydrogen density of $\sim 5
\times 10^4$ cm$^{-3}$ in the shell
surrounding the cavity in the molecular emission. Fuente et al. (1993)
mapped NGC 7023 at higher angular resolution (11$''$ to 26$''$ half-power
beamwidth) in a number of molecular millimeter lines,
including HCO$^+$, and found evidence for gas densities of up to $\sim
10^5$cm$^{-3}$. High resolution
maps of NGC 7023 in HCO$^+$ and HI, in particular of the molecular rim
surrounding the cavity containing atomic gas, by Fuente et al. (1996)
have found four high-density molecular filaments with velocities from
1.9 to 4.0 km s$^{-1}$. These HCO$^+$ filaments match the filaments shown in
optical polarization color images (Watkin, Gledhill, \& Scarrott 1991),
2.1 $\mu$m broadband images (Sellgren, Werner, \& Dinerstein 1992),
narrowband images of $1-0$ S(1) \h2 emission (LeMaire et al. 1996), and
mid-IR images (Cesarsky et al. 1996).
Fuente et al. (1996) estimate the density of these filaments to be
$\sim 10^5$ cm$^{-3}$ from the fractional abundance of HCO$^+$.
This is a factor of $10 - 25$ times greater than the densities reported 
by Chokshi et al. (1988) and Rogers et al. (1995) at lower spatial resolution.

\section{Observations}

The observations of NGC 7023 discussed in this paper were made on
1993 July 5 with (FWHM) source sizes of $\sim 1\farcs4 \times 0\farcs9$ 
(including seeing and all instrumental effects) at the University
of Hawaii's 2.2-m telescope on Mauna Kea with KSPEC (Hodapp et al.  1994).
This instrument provides medium
resolution, cross-dispersed spectra from 1 to 2.5 $\mu$m with an $0\farcs6
\times 6\farcs5$ slit oriented NS. At $K$, our spectral resolution was 30 \AA\
($R = \lambda/\Delta \lambda = 730$)
and the dispersion was 20 \AA\ pixel$^{-1}$, leading to
undersampled spectra. At $H$, our spectral resolution was 23 \AA\ ($R = 720$)
and the dispersion was 15 \AA\ pixel$^{-1}$.
The instrument makes use of two $256
\times 256$ NICMOS3 arrays, one for spectroscopic information and the other to
provide imaging of the region around the slit for tracking purposes.

Two different positions in the nebula were observed, both centered on
relatively bright regions. The first is located
40$''$ W, 34$''$ N (hereafter Position 1) of the central star,
HD 200775; the second is on a somewhat fainter filament 22$''$ W, 66$''$ S
(hereafter Position 2) of the central star.
Three 150s integrations were taken of each position and summed together.
The offset induced in the slit position by sky chopping 300$''$ N
between integrations was found to be negligibly small compared to the
width of the filaments ($\gsim 6''$). Spectra of three A0 stars, 
SAO 19255, HR 8300, and the photometric standard HD 203856 (Elias et al.
1982), were then taken for use in flux calibration and atmospheric
correction. Using our spectra and narrowband images obtained by A.
Quillen, we estimate the absolute uncertainty in our flux calibration to
be a factor of 2.4 ($1\sigma$).

Figures 1 and 2 show the final reduced spectra.
The data were reduced using the OSU implementation of VISTA (Stover 1988)
by first straightening the spectra along the slit using the centroid of the
object and then summing over a 5$''$ wide portion of the $J, H,$ and $K$ bands.
The spectra were wavelength calibrated with an Argon lamp, resulting in
a calibration to vacuum wavelengths (e.g. Black \& van Dishoeck 1987).

The line strengths of the reduced and extracted (but not unreddened) spectra
were measured using the OSU LINER package. We achieved a signal-to-noise
approaching 20 for the strongest lines and 10 for the weaker lines (see
Table 1) in the $K$ Band, where this instrument is most sensitive. Our
relative lin intensities are therefore better determined than the
absolute intensities. Lower
quality detections were made in $H$ due to a combination of lower sensitivity,
blending of the \h2 lines, and difficulty subtracting the
night-sky OH airglow emission lines. We report no clear detections of
\h2 lines in our $J$ band spectra due to both
strong atmospheric effects and some overlap of this band with higher
orders in the spectrograph.

An estimation of the extinction towards HD 200775, the exciting star of NGC
7023, is complicated by the fact that HD 200775 is a Herbig Ae/Be star
and thus may have unusual colors and/or circumstellar dust.
Values of $R_{V}$ for HD 200775 range from 3.1 (Witt \& Cottrell 1980) to
6.25 (Aiello et al. 1988). For this analysis, we have adopted the $R_{V} = 5$
reddening law of Mathis (1990) as an intermediate value and have used
$E(B - V)  = 0.44$ (Witt \& Cottrell 1980). All flux measurements quoted in
subsequent analysis have been dereddened using this extinction law.
The effects of extinction on our results is discussed in \S 6.

\section{Rotational and Vibrational Temperatures} \label{RVT}

A clear indicator of the presence of UV-pumped fluorescence is a difference 
in the rotational and vibrational temperatures. The rotational
(or vibrational) temperature is measured from the slope of a line
passing through data with different rotational (or vibrational) upper levels but
the same vibrational (or rotational) upper levels on diagrams of the
ln($N_{u}/g_{u})-T_{u}$ plane, respectively (e.g. Figures 3 and 4).
Thermal excitation, with an OPR of 3, will give similar rotational and
vibrational temperatures, while fluorescence is characterized by a high
vibrational temperature and a low rotational temperature (e.g.
Tanaka et al. 1989).

Figures 3 and 4 show that both positions observed in NGC 7023 are
characterized by high vibrational temperatures ($\sim$ 5000 K) and low
rotational temperatures ($\sim$ 1000 K).
Our calculated rotation temperature is higher than the value derived by
LeMaire et al. (1996), $\sim$ 500 K, but is still within the
uncertainties of
their $1-0$ S(2) data. Tables 2 and 3 show the calculated vibrational and
rotational temperatures between neighboring ortho and para states
for several of the higher signal-to-noise transitions.
This difference between rotational and vibrational
temperatures was predicted by
Black \& Dalgarno (1976) and Black \& van Dishoeck (1987)
and is a definite indicator of the prescence of fluorescent
emission (see Burton (1992) for a review and in particular Figure 4 of
that paper).

\section{Ortho-to-Para Ratio} \label{OP}

The ortho-to-para ratio (OPR) of molecular hydrogen is the ratio of the total
column density of ortho-\h2 (odd $J$) to para-\h2 (even $J$),
divided by their respective statistical weights. The statistical
weight of the ortho and para states, $g_{u}$, is the product of the
rotational ($g_{J} = 2J + 1$) and spin ($g_{s} = 1$ for even $J$ and 3
for odd $J$) degeneracies.

When \h2 is formed on the surface of grains in molecular clouds,
it is usually assumed to initially have an OPR of 3 
(see e.g.  Spitzer \& Zweibel 1974).
After formation, the OPR can evolve with time, either while still on the
grain surface (Tielens \& Allamandola 1987), or after ejection into the
gas phase, depending upon the conditions inside the cloud (Dalgarno et
al. 1973; Flower \& Watt 1984; Hasegawa et al. 1989; Tanaka et al. 1989;
Chrysostomou et al. 1993). Exchange reactions with atomic hydrogen and
H$^+$ can alter the OPR (Dalgarno, Black \& Weisheit 1973; Flower \&
Watt 1984), though in regions with low H$^+$ number density, such as NGC
7023, reactions involving atomic hydrogen will dominate and the rate of
this reaction will strongly depend on the temperature (Takayanagi et al.
1987). For high temperatures, the OPR will be 3, while much lower values
may occur for T $\lsim 200$ K (Burton et al. 1992). Furthermore, if 
a photodissociation front is exposing new, cool material to the warm UV
field faster than these exchange reactions can drive the OPR towards 3,
a low OPR could be maintained (Chrysostomou et al. 1993). The OPR thus
may reflect the formation conditions or may indicate that exchange
reactions are taking place. Values in the range $1 - 1.8$ have been
observed in fluorescent \h2 sources (Tanaka et al. 1989; Ramsay et al.
1993; Chrysostomou et al. 1993; Hora \& Latter 1996). 

In Figures 3 and 4 we show ln($N_{u}/g_{u}$) vs. $T_{u}$ derived from
our dereddened spectra of NGC 7023. The excitation temperature of the
upper level (Dabrowski 1984) is equal to the energy of the upper level divided 
by Boltzmann's constant ($T_{u} = E_{u}/k$), and $N_{u}$ is the column
density of the upper level. Based upon an examination of such a
plot alone, it is possible to determine a great deal about the nature of
the PDR. We plot different symbols for ortho (filled points) and para
(open points) \h2 in Figures 3 and 4 to illustrate the effect of the
OPR. When ortho and para measurements for a given vibrational state fall
on the same line, this corresponds to an OPR of 3 and is
indicative of thermal processes. If the derived level populations of ortho 
and para states for a given
vibrational state separate vertically in the ln($N_{u}/g_{u})-T_{u}$
plane, the OPR is different from 3, as is often observed for UV
fluorescence. 

Following the method outlined by Tanaka et al. (1989), we computed the
vertical separation between the $J = 3$ and 4 levels by a least-squares
fit through the neighboring para or ortho pair of points on the diagram,
respectively, for $v = 1$. We did not use $v = 2$ due to telluric
contamination in the $2-1$ S(2) line (Everett 1997). We obtained an OPR
of $2.5 \pm 0.3$ for Position 1 and an OPR of $2.4 \pm 0.5$ for Position
2. Both of these values are intermediate between the values for
thermalized gas (3) and those observed for fluorescent emission
($1 - 1.8$), where the difference from 3 is primarily driven by the $J =
2$ level. Though there is clear evidence that significant fluorescent
emission is occurring in NGC 7023 (see, e.g. \S 3), the fact that our
values for the OPR are higher than has been previously observed in
fluorescent \h2 sources may indicate that additional thermal processes,
such as contributions from shocks, are contributing to the chemistry of
this PDR. This calculation of the OPR for NGC 7023 shows how the OPR can
be a useful tool to investigate the nature of PDRs and that the
interpretation of the OPR ratio can be complicated by the presence of
shocks, which can mimic the behavior of warm gas. 

\section{Models} \label{Models}

Based on the previous observations of NGC 7023 outlined in the introduction,
the incident UV flux has a possible range of $G_{0} = 10^{3}$ to 10$^{4}$,
while the density of the surrounding material has a possible range of 10$^{2}$
cm$^{-3}$ in the more diffuse regions to 10$^{5}$ cm$^{-3}$ in the denser
regions near HD 200775. We can also use our observed $1-0$ S(1) line
intensity as an additional constraint in modeling the PDR in the two
molecular filaments we investigate here by eliminating models which
predict $1-0$ S(1) intensities differing by more than $3\sigma$ (or a
factor of 14) from our observed intensities. 

If NGC 7023 is composed of low-density, UV-illuminated clouds
($n < 10^{4}$ cm$^{-3}$), it will show fluorescent \h2 emission (Black \&
van Dishoeck 1987).  In UV-illuminated clouds with densities above the
critical density, $n > 10^{5}$ cm$^{-3}$, collisional deexcitation
occurs faster than radiative decay and this causes some \h2 line
ratios to approach values for thermal excitation (Sternberg \& Dalgarno
1989;  Burton et al. 1990; Draine \& Bertoldi 1996).
Draine \& Bertoldi (1996) have found a density of
$10^5$ cm$^{-3}$ to provide the best agreement with NGC 2023, showing
that collisional deexcitation dominates over the radiative fluorescent
cascade in this case.
If there is a bipolar outflow in NGC 7023, this could heat the gas by
shocks, giving rise to a thermal component of \h2 emission.

To determine the nature of the molecular filaments in NGC 7023, we have
compared our dereddened flux measurements to the PDR models of Draine \&
Bertoldi (1996), which include both radiative transfer and the best
available collisional data. The basic characteristics of all the
best-fitting models described in the text are summarized in Table 4.
Draine \& Bertoldi (1996) present 
models ranging from $n = 10^{2}$ to $10^{6}$ cm$^{-3}$ and in UV flux from
$G_{0} = 1$ to $10^{5}$. In addition to single model fits, we have also
mixed the models of Draine \& Bertoldi (1996) with the thermal models of
Black \& van Dishoeck (1987). These thermal models, referred to as S1
and S2, represent shock-excited gas at temperatures of 1000 and 2000 K,
respectively. 

In order to compare our data with these models, 
we computed the value of ln(N$_{u}$/g$_{u}$) normalized to the
$1-0$ S(1) line for each model and our data. We then calculated $\chisq$
to determine the goodness of fit. Our $\chisq$ parameter was calculated
by varying only one free parameter, the overall intensity. The
uncertainty used in calculating $\chisq$ only reflects uncertainties in
the relative line intensities. We considered any model with 
$\chisq \leq \chisq_{min} + 1$, where $\chisq_{min}$ corresponds to the
best-fitting model for that filament, to be a reasonable fit (Bevington
\& Robinson 1992). We note that our $\chisq$ value indicates which
models give the best overall agreement with our observations, but does
not indicate which regions of the ln($N_{u}/g_{u})-T_{u}$ plane were in good or
poor agreement. The 10 lines used in our analysis were $v = 1$, $J = 2 - 5$;
$v = 2$, $J = 2$, 3, and 5; and $v = 3$, $J = 3 - 5$. As mentioned above,
the $2-1$ S(2) line was excluded as it lies near a contaminating telluric
feature (Everett 1997). In addition, although the $3-2$ S(2) line
intensity was not above the $3\sigma$ detection limit at Position 2, we
used the measured flux and associated error to include this line in the
$\chisq$ calculation. The results of these calculations are listed in
Table 4 along with the physical characteristics of the best-fitting
models. 

We next considered mixtures of models. To mix two models together, we
normalized the flux predictions for each separate model to the $1-0$ S(1)
line and multiplied each model by its percentage contribution to the
final mixture. This mixture was then converted to ln($N_{u}/g_{u}$)
format and $\chisq$ was calculated as described above. When mixing
two models, we stepped from 0\% to 99\% contribution of the second model
in 1\% increments. We explored fluorescent models, at a range of densities,
mixed with thermal models. We also examined low-density fluorescent models
mixed with high-density fluorescent models. Table 5 lists our results for
mixtures of
fluorescent/collisional models with a thermal component. In our analysis, 
we excluded all models with a $1-0$ S(1) intensity less than 
$\sim 2 \times 10^{-5}$ erg s$^{-1}$ cm$^{-2}$ sr$^{-1}$ based on our 
flux calibration.
Table 6 lists our results for mixtures of the models of
Draine \& Bertoldi (1996) with their high-density ($n = 10^{6}$ cm$^{-3}$)
model (Qm3o). Mixtures of this model with lower-density models produced the
best fits to our data of all of our high + low-density combinations.

\section{Discussion} \label{Discussion}

For Position 1, the model comparisons presented in Tables 4, 5, and 6
show that a mixture of a warm component with fluorescent emission fits
somewhat, but not significantly, better than pure fluorescent emission.
The improvement in the fit is $\Delta\chisq = 1.4$, where $\Delta\chisq
= 1$ corresponds to $1\sigma$ (Bevington \& Robinson 1992). The warm
component which improves the fit is equally well characterized by a high
density fluorescent model, with $n = 10^6$ cm$^{-3}$ and $G_0 = 10^4$, or
by a thermal model with $T = 1000 - 2000$ K. The best-fitting mixed
models have $n = 10^4 - 10^5$ cm$^{-3}$ and $G_0 = 1000 - 5000$ for
the low-density fluorescent component. The best-fit pure fluorescent
models, not surprisingly, have values that are intermediate between the
values for the low-density and high-density components of the best-fit
mixed models: $n = 10^5 - 10^6$ cm$^{-3}$ and $G_0 = 10^3 - 10^4$. 

For Position 2, the data were equally well fit by pure fluorescent
models or by mixtures of a warm component with fluorescent emission
($\Delta\chisq \leq 1$). In all cases, the best-fit models have 
$n = 10^4 - 10^5$ cm$^{-3}$ and $G_0 = 10^3$. 

The best-fit models as determined by the $\chisq$ results in Tables 4 through
6 are slightly dependent upon the extinction. As the
rotational and vibrational quantum numbers, $v$ and $J$, decrease with
increasing wavelength (see e.g. Table 1), a larger value of the extinction
increases the intensity of the lower (shorter wavelength) transitions
over the upper (longer wavelength) transitions. Thus an underestimation
of the extinction can mimic the effect of adding a higher-temperature thermal
component or a higher-density collisional component to the excitation.
We found, however, that assuming either no extinction or twice our adopted
value ($2 \times E(B-V) = 0.88$) changes our results in Tables 4, 5, \&
6 by $\Delta\chisq < 1$ (less than $1\sigma$). 

As mentioned in the previous section, our $\chisq$ parameter only
tells us approximately which models provide the best agreement to the
data and no information on the behavior of individual components of the
spectrum. In addition, the $\chisq$ parameter only includes the
uncertainties in our relative line intensity measurements, and does not
reflect possible uncertainties in the models or the effects of a more
complex geometry than were employed in the model calculations. 
One way of extracting information about areas of specific
disagreement is to test the data and models for lines ratios most strongly
affected by a warm component. 

Figure 5 shows a plot of the $2-1$ S(1)/$1-0$ S(1) vs. $2-1$ S(3)/$1-0$ S(3) 
line ratios and includes both our data and representative models from Table 4.
Figure 5 also shows the behavior of these line ratios for the $n = 10^4$
cm$^{-3}$, $G_0 = 10^3$ model (Hw3o) of Draine \& Bertoldi (1996) as it
is mixed with greater percentages of two thermal models 
(Black \& van Dishoeck 1987). These lines are well-suited to testing the
contribution of a thermal component to a fluorescent model as the
addition of a thermal component to a fluorescent model tends to enhance
the $v = 1$ odd $J$ levels in the spectrum as 
it pushes the OPR towards 3, but has less effect on the $v = 2$ odd $J$ 
levels, as the thermal component contributes most to
the lower-lying levels. The rotational temperatures for the $v = 1$ levels
(except for $J = 4 - 2$)
are also higher than those of the $v = 2$ and 3 levels, providing
further evidence of the presence of a second, possibly thermal,
component. This figure provides some tantalizing evidence that there is
a thermal component to the emission that further investigation with
higher signal-to-noise data or a greater number of lines could resolve.

Figure 5 also shows that none of the pure fluorescent models, and none
of the mixtures of fluorescence with a thermal component or a higher-density 
component, can explain the very strong $1-0$ S(3) emission we
observe. It is possible that this line is blended with some other
unknown line, or that we have underestimated the uncertainties in the 
atmospheric correction at this wavelength. As a check, we compared all
the models in Tables 4, 5, and 6 to our observations with the $1-0$ S(3)
line excluded. The minimum $\chisq$ values improve significantly
($\chisq_{min} = 1.3$ for Position 1 and 2.9 for Position 2), but our
overall conclusions concerning the density and excitation in NGC 7023
are unchanged. 

\section{Conclusion}

Analysis of the near-infrared lines of molecular hydrogen in NGC 7023 has
shown that the best-fit pure fluorescent models for Position 1 have $n =
10^5 - 10^6$ cm$^{-3}$ and $G_0 = 10^3 - 10^4$. The model fits are
somewhat improved ($1.4\sigma$) by a mixture of models involving a
fluorescent model with $n = 10^4 - 10^5$ cm$^{-3}$ and 
$G_0 = 1000 - 5000$ and a warm component. This warm component could be
either a higher-density, fluorescent model ($n = 10^6$ cm$^{-3}$, 
$G_0 = 10^4$), perhaps due to dense clumps in the PDR, or a model
reflecting a thermal contribution, possibly due to shocks. In Position
2, the best-fitting models have $n = 10^4$ cm$^{-3}$ and $G_0 = 10^3$,
with our data equally well fit by fluorescence with or without a warm
component. Our derived densities,  $n = 10^4 - 10^5$ cm$^{-3}$ 
Position 1 and $10^4$ cm$^{-3}$ in Position 2, are in good agreement with
previously determined densities of $4 \times 10^3 - 10^5$ cm$^{-3}$.
The suggestion of a warm component in Position 1 may either point to
high-density clumps ($10^6$ cm$^{-3}$) or suggest shocks due to a
bipolar outflow. 

\vskip 24pt

\acknowledgments

We thank Michael Burton and Amiel Sternberg for very generously sending us
more detailed model predictions based on their published work. We also
appreciate Bruce Draine making his model predictions available electronically.
We are grateful to Darren DePoy, Mark Everett, and Rick Pogge for useful
comments on the manuscript. Special thanks are due to the referee, Xander 
Tielens, who also provided many helpful comments that greatly improved this 
presentation. We would also like 
to thank Alice Quillen for obtaining narrowband images of NGC 7023 for us. 
We appreciate the guidance 
of Adolf Witt, who suggested positions to observe in NGC 7023 based on his 
unpublished \h2 imaging. Partial financial support was provided 
by the Alfred P. Sloan Foundation (KS) and the Astronomy Department at OSU (PM).

\clearpage

\begin{center}
\begin{tabular}{cccccc}
\multicolumn{6}{c}{{\bf TABLE 1}}\\[12pt]
\multicolumn{6}{c}{{\bf Normalized Line Intensities}}\\[12pt]
\hline
\hline
\multicolumn{1}{c}{Transition} &
\multicolumn{1}{c}{Wavelength($\mu$m)} &
\multicolumn{2}{c}{Position 1} &
\multicolumn{2}{c}{Position 2} \\
	 &	   & Intensity & $\sigma_{I}$ & Intensity & $\sigma_{I}$ \\
\hline
$1-0$ S(7) & 1.7475  & 0.199 & $\pm$ 0.040    & 0.250 & $\pm$ 0.050 \\
$1-0$ S(6)\tablenotemark{1} & 1.7876  & $<$0.21 &        & $<$0.18 &     \\
$2-1$ S(5) & 1.9443  & 0.299 & $\pm$ 0.037    & 0.225 & $\pm$ 0.029 \\ 
$1-0$ S(3) & 1.9570  & 1.237 & $\pm$ 0.085    & 1.103 & $\pm$ 0.086 \\
$2-1$ S(4)\tablenotemark{1} & 2.0035  & $<$0.10&       & 0.093 & $\pm$ 0.030 \\
$7-5$ O(5) & 2.0215  & 0.063 & $\pm$ 0.020    & 0.125 & $\pm$ 0.030 \\
$1-0$ S(2) & 2.0332  & 0.377 & $\pm$ 0.025    & 0.400 & $\pm$ 0.038 \\
$8-6$ O(3) & 2.0412  & 0.071 & $\pm$ 0.020    & $<$0.09 &     \\
$3-2$ S(5) & 2.0650  & 0.111 & $\pm$ 0.025    & $<$0.110&     \\
$2-1$ S(3) & 2.0729  & 0.293 & $\pm$ 0.021    & 0.249 & $\pm$ 0.030 \\
$1-0$ S(1) & 2.1213  & 1.000 & $\pm$ 0.039    & 1.000 & $\pm$ 0.043 \\
$3-2$ S(4) & 2.1274  & $<$0.09 &        & $<$0.09 &     \\
$2-1$ S(2)\tablenotemark{1} & 2.1536  & 0.161 & $\pm$ 0.020    & 0.148 & $\pm$ 0.025 \\
$3-2$ S(3) & 2.2008  & 0.147 & $\pm$ 0.018    & 0.109 & $\pm$ 0.030 \\
$1-0$ S(0) & 2.2227  & 0.383 & $\pm$ 0.025    & 0.463 & $\pm$ 0.049 \\
$2-1$ S(1) & 2.2471  & 0.325 & $\pm$ 0.028    & 0.446 & $\pm$ 0.028 \\
$3-2$ S(2) & 2.2864  & 0.081 & $\pm$ 0.017    & 0.060\tablenotemark{2} & $\pm$ 0.030 \\
$2-1$ S(0) & 2.3550  & 0.186 & $\pm$ 0.018    & 0.286 & $\pm$ 0.026 \\
$3-2$ S(1) & 2.3858  & 0.245 & $\pm$ 0.025    & 0.268 & $\pm$ 0.028 \\ 
$1-0$ Q(1) & 2.4059  & 1.800 & $\pm$ 0.086    & 1.651 & $\pm$ 0.074 \\
$1-0$ Q(2) & 2.4128  & 0.661 & $\pm$ 0.051    & 0.708 & $\pm$ 0.041 \\
\hline
\end{tabular}
\end{center}

\noindent
$^{1}$ This line may suffer contamination by telluric absorption. 

\noindent
$^{2}$ Non-detection: measured line intensity is less than the $3\sigma$ limit.

\noindent 
Table 1: Normalized line intensities for Positions 1 and 2 in NGC 
7023. Position 1 is 40$''$ W, 34$''$ N of HD 200775 and Position 2 is 
22$''$ W, 66$''$ S of HD 200775. Columns 1 and 2 list the molecular transitions 
and vacuum wavelengths (after Black \& van Dishoeck 1987) for the emission 
lines. Columns 3 and 4 list the intensities and 1$\sigma$ errors for 
lines at Position 1. The intensities have been normalized with respect
to the $1-0$ S(1) intensity of $5.43 \pm 1.64 \times10^{-4}$ ergs cm$^{-2}$ 
s$^{-1}$ sr$^{-1}$. Columns 5 and 6 list the intensities and 1$\sigma$ errors 
for emission lines at Position 2. These intensities have been normalized with 
respect to the $1-0$ S(1) intensity of $2.58 \pm 0.78 \times10^{-4}$ ergs 
cm$^{-2}$ s$^{-1}$ sr$^{-1}$. The uncertainties in the $1-0$ S(1) intensity 
include loss at the slit but do not reflect the uncertainty in the
overall flux calibration (see \S 2). All upper limits are 3$\sigma$ limits.

\clearpage
\begin{center}
\begin{tabular}{ccccccc}
\multicolumn{7}{c}{{\bf TABLE 2}}\\[12pt]
\multicolumn{7}{c}{{\bf Rotation Temperature}}\\
\hline
\hline
\multicolumn{1}{c}{Transition}	&
\multicolumn{1}{c}{Level 1} &
\multicolumn{1}{c}{Level 2} &
\multicolumn{2}{c}{Position 1} &
\multicolumn{2}{c}{Position 2} \\
         &          &	& T$_{rot}$ & $\sigma_{T}$ & T$_{rot}$ & $\sigma_{T}$ \\
\hline
$J = 4-2$ &$1-0$ S(0) & $1-0$ S(2) &  970 & $\pm$  80 & 870  & $\pm$ 130 \\
$J = 5-3$ &$1-0$ S(1) & $1-0$ S(3) & 2750 & $\pm$ 450 & 2240 & $\pm$ 500 \\
$J = 9-5$ &$1-0$ S(3) & $1-0$ S(7) & 2080 & $\pm$ 210 & 2480 & $\pm$ 330 \\
$J = 5-3$ &$2-1$ S(1) & $2-1$ S(3) & 1710 & $\pm$ 330 & 1060 & $\pm$ 120 \\
$J = 5-3$ &$3-2$ S(1) & $3-2$ S(3) & 1120 & $\pm$ 160 & 830  & $\pm$ 160 \\
\hline
\end{tabular}
\end{center}

\noindent 
Table 2: Rotation temperatures for \h2 emission in NGC 7023 
calculated from the relative line
strengths for the listed transitions with associated $1\sigma$ 
uncertainties.

\clearpage
\begin{center}
\begin{tabular}{ccccccc}
\multicolumn{7}{c}{{\bf TABLE 3}}\\[12pt]
\multicolumn{7}{c}{{\bf Vibration Temperature}}\\
\hline
\hline
\multicolumn{1}{c}{Transition} &
\multicolumn{1}{c}{Level 1} &
\multicolumn{1}{c}{Level 2} &
\multicolumn{2}{c}{Position 1} &
\multicolumn{2}{c}{Position 2} \\
         &          &	& T$_{vib}$ & $\sigma_{T}$ & T$_{vib}$ & $\sigma_{T}$ \\
\hline
$v = 2-1$&$2-1$ S(0) & $1-0$ S(0) & 5410 & $\pm$ 610  & 7040 & $\pm$ 1660 \\
$v = 2-1$&$2-1$ S(1) & $1-0$ S(1) & 3920 & $\pm$ 260  & 5030 & $\pm$  430 \\
$v = 2-1$&$2-1$ S(3) & $1-0$ S(3) & 3250 & $\pm$ 280  & 3160 & $\pm$  320 \\
$v = 3-2$&$3-2$ S(3) & $2-1$ S(3) & 8580 & $\pm$ 2490 & 7010 & $\pm$ 2840 \\
\hline
\end{tabular}
\end{center}

\noindent 
Table 3: Vibration temperatures for \h2 emission in NGC 7023 
calculated from the relative line
strengths for the listed transitions with associated $1\sigma$
uncertainties. 

\clearpage

\begin{center}
\begin{tabular}{lcccccc}
\multicolumn{7}{c}{{\bf TABLE 4}}\\[12pt]
\multicolumn{7}{c}{{\bf Model Characteristics}}\\
\hline
\hline
\multicolumn{1}{c}{Model} &
\multicolumn{2}{c}{$\chisq$} &
\multicolumn{1}{c}{$n_{H}$} &
\multicolumn{1}{c}{$G_{0}$} &
\multicolumn{1}{c}{T$_{gas}$} &
\multicolumn{1}{c}{I$_{1-0 S(1)}$} \\
	  & Position 1 & Position 2 & cm$^{-3}$	& (Habings)	& (K) & erg s$^{-1}$ cm$^{-2}$ sr$^{-1}$ \\
\hline
Hw3o	  &   12.   &{\bf 4.4}	& $10^{4}$	& $10^{3}$     	& 500 & $1.85 \times 10^{-5}$\\
Hh3o	  &   8.0   &{\bf 5.1}	& $10^{4}$	& $10^{3}$     	& 1000 & $2.20 \times 10^{-5}$\\
Lm3o	  &   7.5   &{\bf 5.2}	& $10^{5}$	& $10^{3}$	& 300 & $2.61 \times 10^{-5}$\\	
Lw3o	  &{\bf 6.2}&   5.5 	& $10^{5}$	& $10^{3}$	& 500 & $2.69 \times 10^{-5}$\\	
Mw3o      &   8.1   &   8.3	& $10^{5}$	& $10^{4}$	& 500 & $9.66 \times 10^{-5}$\\
Mh3o	  &{\bf 6.7}&   9.4     & $10^{5}$	& $10^{4}$	& 1000 & $1.17 \times 10^{-4}$\\
Qm3o	  &{\bf 6.4}&   11.	& $10^{6}$ 	& $10^{4}$	& 300 & $1.32 \times 10^{-4}$\\
Qw3o      &   9.2   &   12.	& $10^{6}$	& $10^{4}$	& 500 & $1.42 \times 10^{-4}$\\
n2023a    &   7.3   &   7.9   	& $10^{5}$	& $5000$	& 500 & $3.29 \times 10^{-4}$ \\
\hline
\end{tabular}
\end{center}

\noindent 
Table 4: Basic characteristics of the best-fitting models of 
Draine \& Bertoldi (1996) compared to Positions 1 and 2 in NGC 7023. 
The $\chisq$ values listed are reduced 
$\chisq$.  All $\chisq$ calculations were performed with the column
densities derived from the dereddened intensities for the 10 \h2 lines
listed in the text. Values of $\chisq$ in bold type are those with
$\chisq \leq \chisq_{min} + 1$, where $\chisq_{min}$ is the minimum
value of $\chisq$ for each position. The density, $n_{H}$,
is the nucleon density, $G_{0}$ is the UV flux in the units of the 
interstellar radiation field (Habing 1968), $T_{gas}$ is the gas
temperature, and $I_{1-0 S(1)}$ is the intensity of the $1-0$ S(1) line.

\clearpage
\begin{center}
\begin{tabular}{lcccccccc}
\multicolumn{9}{c}{{\bf TABLE 5}}\\[12pt]
\multicolumn{9}{c}{{\bf Best Fit Results for Mixes with a Thermal Component}}\\[12pt]
\hline
\hline
\multicolumn{1}{l}{Models} &
\multicolumn{2}{c}{Position 1 + S1} &
\multicolumn{2}{c}{Position 1 + S2} &
\multicolumn{2}{c}{Position 2 + S1} &
\multicolumn{2}{c}{Position 2 + S2} \\
 & $\chisq$ & \% S1 & $\chisq$ & \% S2 & $\chisq$ & \% S1 & $\chisq$ & \% S2 \\
\hline
Hw3o & 8.7  & 28  &{\bf 5.2}& 35  &{\bf 3.9}& 16 &{\bf 3.6}& 18 \\
Hh3o & 6.8  & 18  &{\bf 5.4}& 25  & 5.1 & 6      & 5.1  & 6  \\
Lm3o & 6.3  & 18  &{\bf 4.9}& 25  & 5.1 & 6      & 5.1  & 6 \\
Lw3o &{\bf 5.6}&13&{\bf 4.8}& 20  & 5.5 & 2      & 5.5  & 0 \\
Mw3o & 7.0  & 18  &    6.4  & 22  & 8.2 & 7      & 8.3  & 4 \\
Mh3o & 6.7  & 0   &    6.7  & 0   & 9.4 & 0      & 9.4  & 0 \\
Qm3o & 6.4  & 0   &    6.4  & 0   & 11. & 0      & 11.  & 0 \\
Qw3o & 9.2  & 0   &    9.2  & 0   & 12. & 0      & 12.  & 0 \\
n2023a& 6.4 & 16  &    6.0  & 19  & 7.8 & 5      & 7.9  & 1 \\
\hline
\end{tabular}
\end{center}

\noindent 
Table 5: The best-fit results from a comparison of our 
observations of Positions 1 and 2 in NGC 7023 to mixtures of 
fluorescent models of stationary PDRs (Draine \& Bertoldi 1996) and 
thermal models (Black \& van Dishoeck 1987). S1 and S2 correspond to 
shock-heated gas at 1000 K and 2000 K, respectively.
Column 1 lists the models described in detail in Table 4. Columns 2 and 3 list
the $\chisq$ values and \% of model S1 in the best-fitting mixture of these 
two models to Position 1. Columns 4 and 5
list the same for mixtures of the models in Column 1 and model S2.
The last four columns repeat columns 2 through 5 but for Position 2.
Values of $\chisq$ in bold type are those with
$\chisq \leq \chisq_{min} + 1$, where $\chisq_{min}$ is the minimum
value of $\chisq$ for each position. 

\clearpage
\begin{center}
\begin{tabular}{lcccc}
\multicolumn{5}{c}{{\bf TABLE 6}}\\[12pt]
\multicolumn{5}{c}{{\bf Mixes of a Low + High Density Component}}\\[12pt]
\hline
\hline
\multicolumn{1}{l}{ Models } &
\multicolumn{2}{c}{ Position 1 + Qm3o } &
\multicolumn{2}{c}{ Position 2 + Qm3o } \\
  	& $\chisq$   & \% Qm3o   & $\chisq$   & \% Qm3o         \\
\hline
Hw3o   & {\bf 4.8}   & 71  & {\bf 4.3} & 12     \\
Hh3o   & {\bf 5.0}   & 62  & {\bf 5.1} & 0      \\
Lm3o   & {\bf 5.1}   & 60  & {\bf 5.2} & 0      \\
Lw3o   & {\bf 5.0}   & 51  & 5.5       & 0      \\
Mw3o   & 5.9         & 71  & 8.3       & 0      \\
Mh3o   & 6.1         & 58  & 9.4       & 0      \\
Qw3o   & 9.2         & 0   & 11.       & 65     \\
n2023a & {\bf 5.8}   & 63  & 7.9       & 0      \\
\hline
\end{tabular}
\end{center}

\noindent 
Table 6: The best-fit results from a comparison of observations
of Positions 1 and 2 in NGC 7023 to mixtures of models from Draine \& 
Bertoldi (1996) with their $n = 10^{6}$ cm$^{-3}$, $G_{0} = 10^{4}$ model
(Qm3o). Column 1 lists the models described in detail in Table 4. 
Columns 2 and 3 list the $\chisq$ values and
\% of model Qm3o in the best-fitting mixture of these models to Position 1.
Columns 4 and 5 list the same for Position 2.  
Values of $\chisq$ in bold type are those with
$\chisq \leq \chisq_{min} + 1$, where $\chisq_{min}$ is the minimum
value of $\chisq$ for each position.

\begin{figure}
\plotfiddle{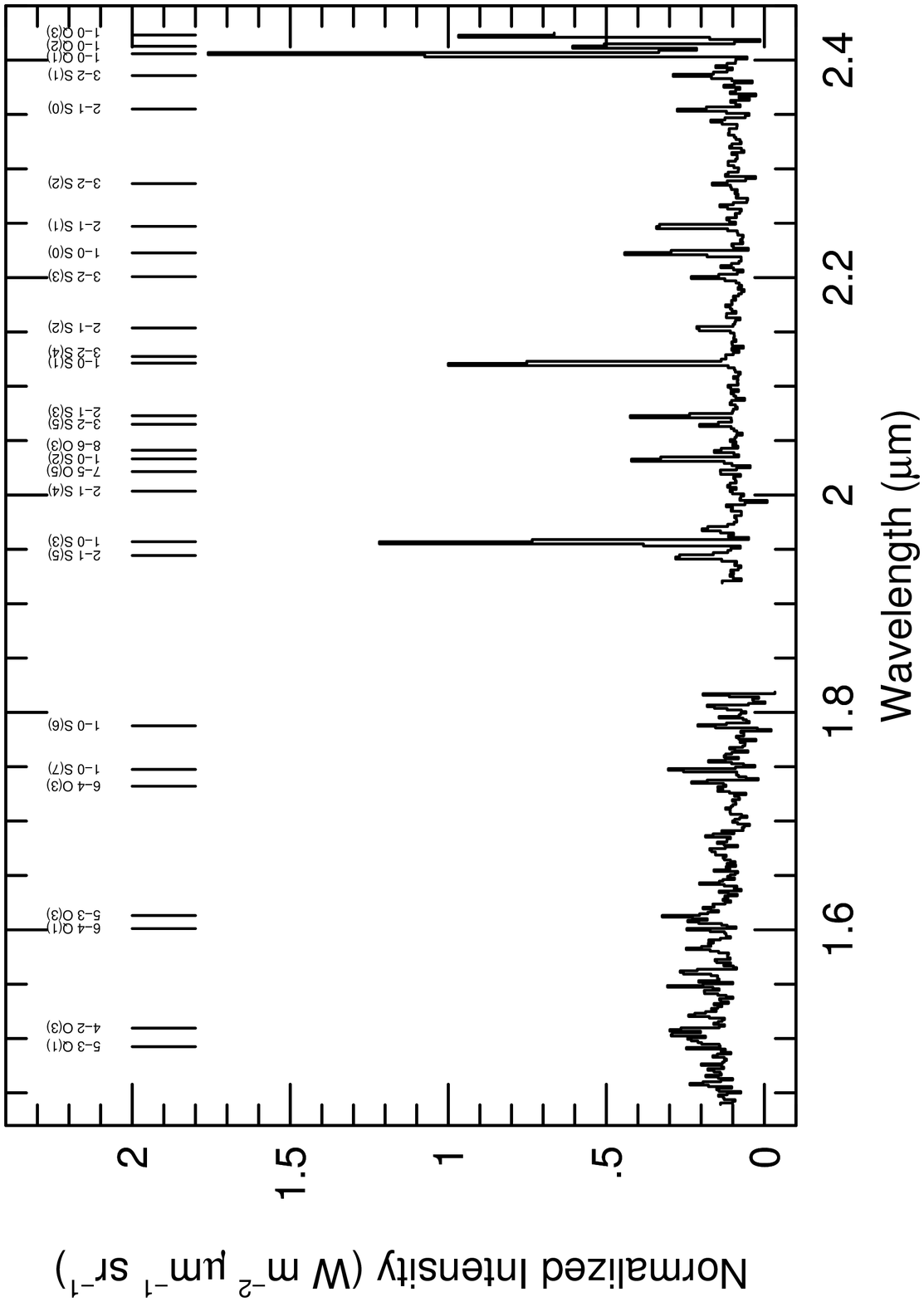}{5.0truein}{0}{70}{70}{-280}{0}
\caption{$H$ and $K$ band spectra of Position 1 in NGC 7023.
Position 1 is located 40$''$ West, 34$''$ North of the central star,
HD 200775. The wavelengths of molecular hydrogen lines have been labeled above
the spectrum. The spectral resolution is $\lambda/\Delta\lambda =
730$ in $K$ and 720 in $H$. The intensity has been normalized to the peak
$1-0$ S(1) intensity. }
\end{figure}

\begin{figure}
\plotfiddle{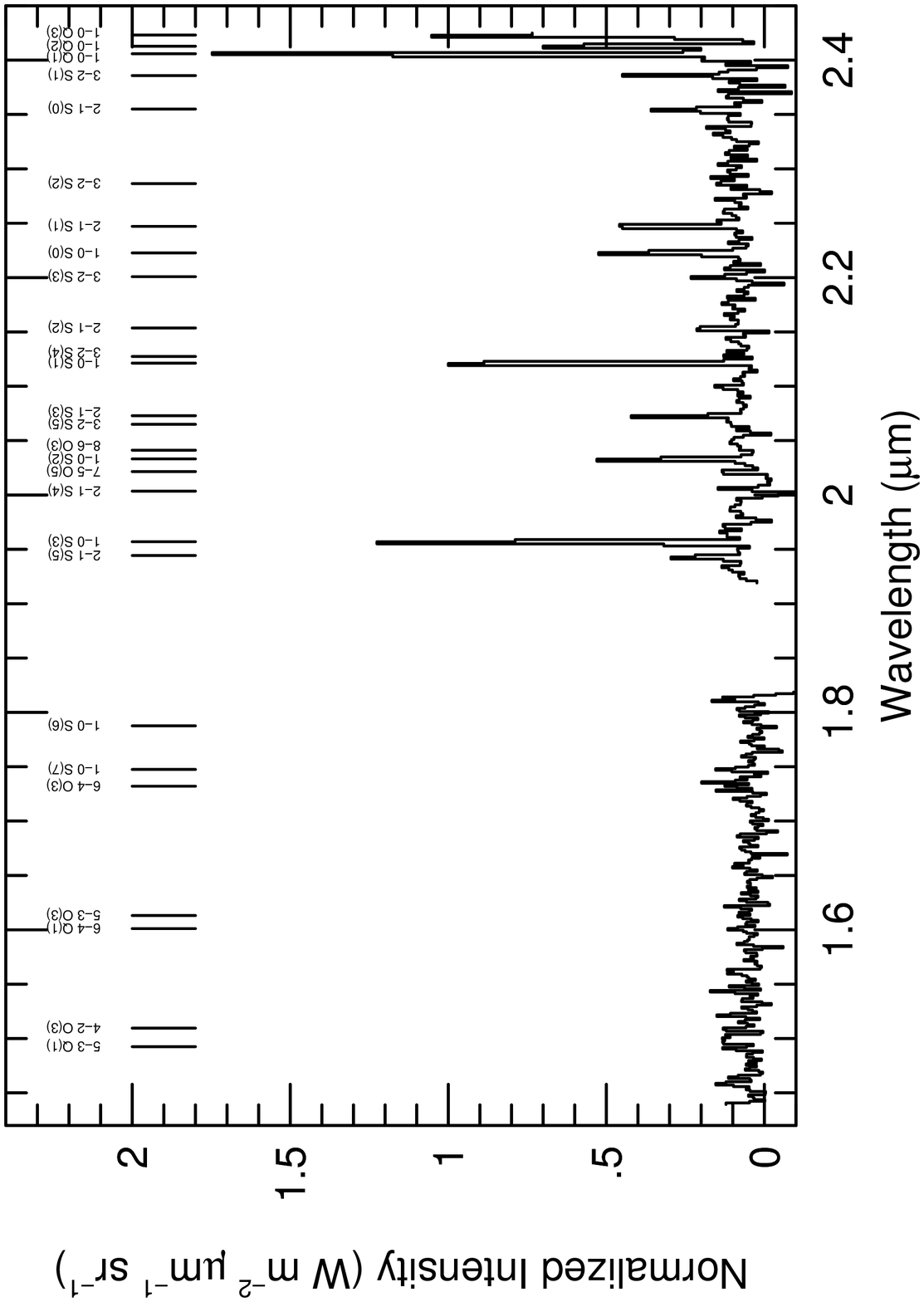}{5.0truein}{0}{70}{70}{-280}{0}
\caption{$H$ and $K$ band spectra of Position 2 in NGC 7023.
Position 2 is located 22$''$ West, 66$''$ South of the central star,
HD 200775. The wavelengths of molecular hydrogen lines have been labeled above
the spectrum as in Figure 1. The spectral resolution is
$\lambda/\Delta\lambda = 730$ in $K$ and 720 in $H$. The intensity has been
normalized to the peak $1-0$ S(1) intensity.}
\end{figure}

\begin{figure}
\plotfiddle{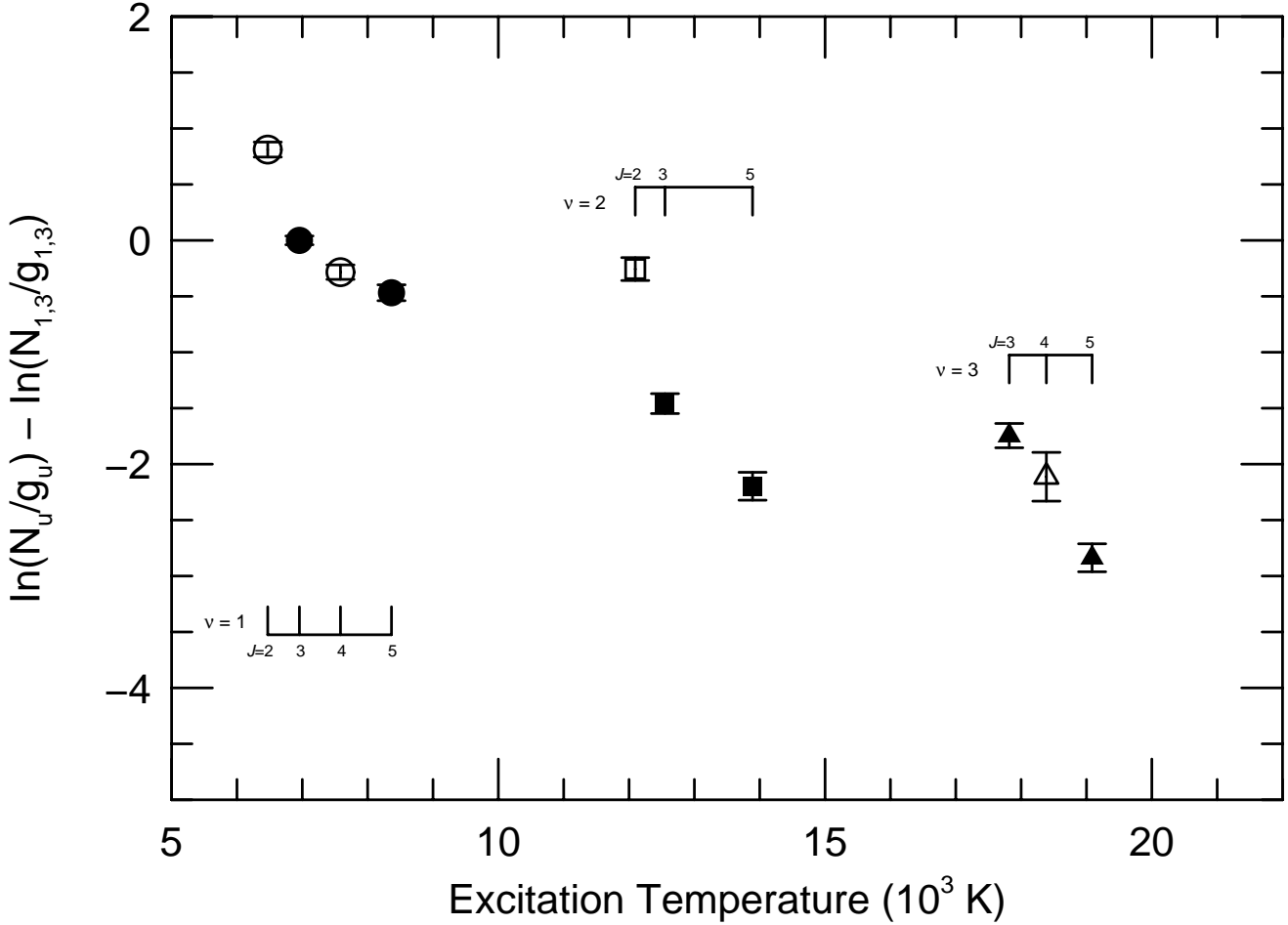}{5.0truein}{0}{70}{70}{-280}{0}
\caption{Diagram of ln($N_{u}/g_{u}$) vs. $T_{u}$ for Position 1
in NGC 7023 for the 10 lines used in our analysis and based upon the data
recorded in Table 1 after dereddening. The $y-$axis shows the log of the column
density of molecular hydrogen in a given state, $N_u$, divided by the
statistical weight, $g_{u}$, and normalized to the $1-0$ S(1) transition
($v = 1, J = 3$). The $x-$axis gives the excitation temperature, $T_{u}$, of
the upper level of a given transition after Dabrowski (1984). Circles denote
transitions from the $v = 1$ vibrational state, squares from $v = 2$, and
triangles from
$v = 3$.  Open symbols represent para (even $J$) rotational transitions,
filled symbols represent ortho (odd $J$) transitions. All errorbars are
$1\sigma$ errors. Upper limits shown are $3\sigma$ upper limits. }
\end{figure}

\begin{figure}
\plotfiddle{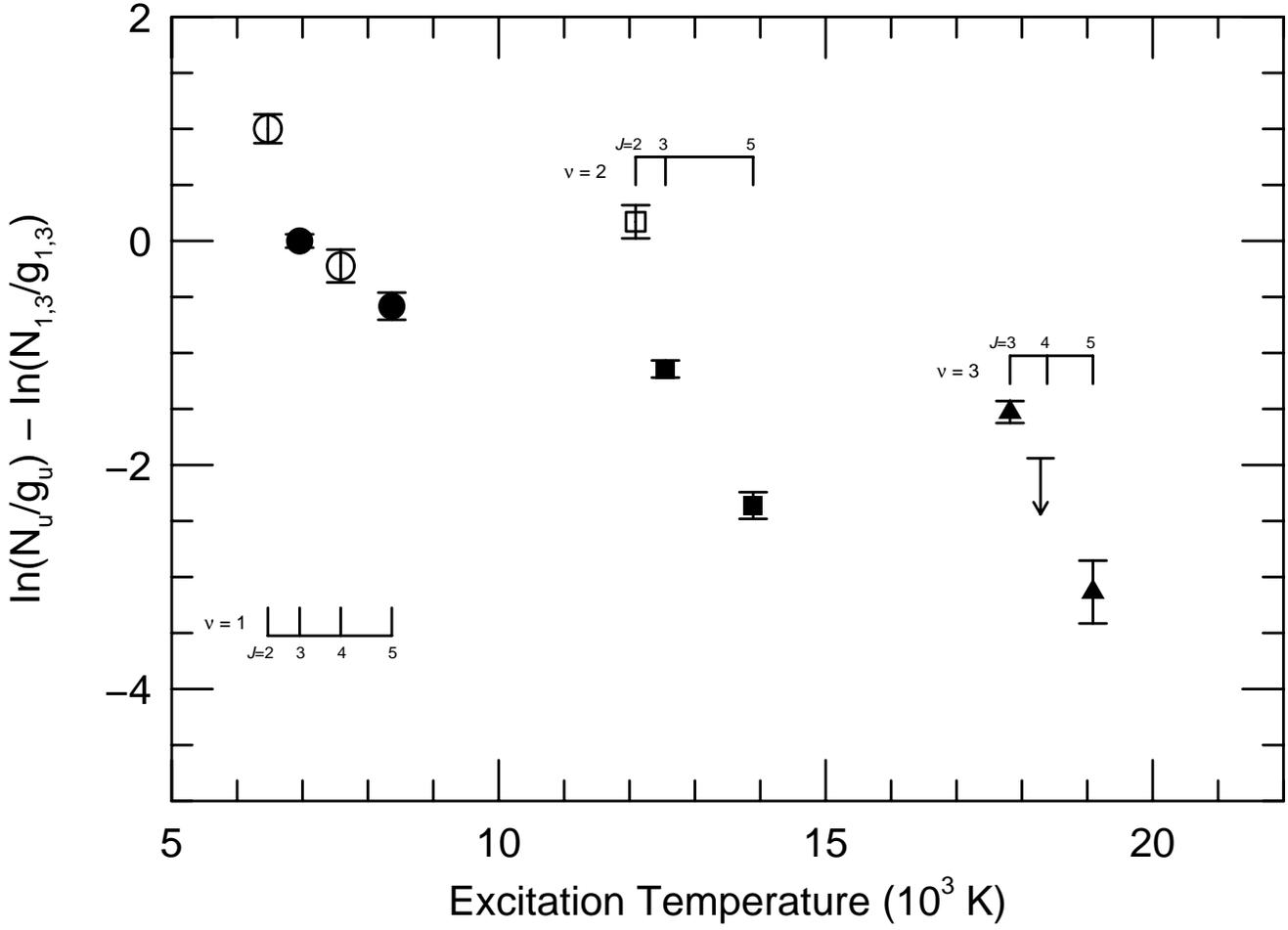}{5.0truein}{0}{70}{70}{-280}{0}
\caption{Same as Figure 3, except for line strengths
measured in Position 2 of NGC 7023.}
\end{figure}

\begin{figure}
\plotfiddle{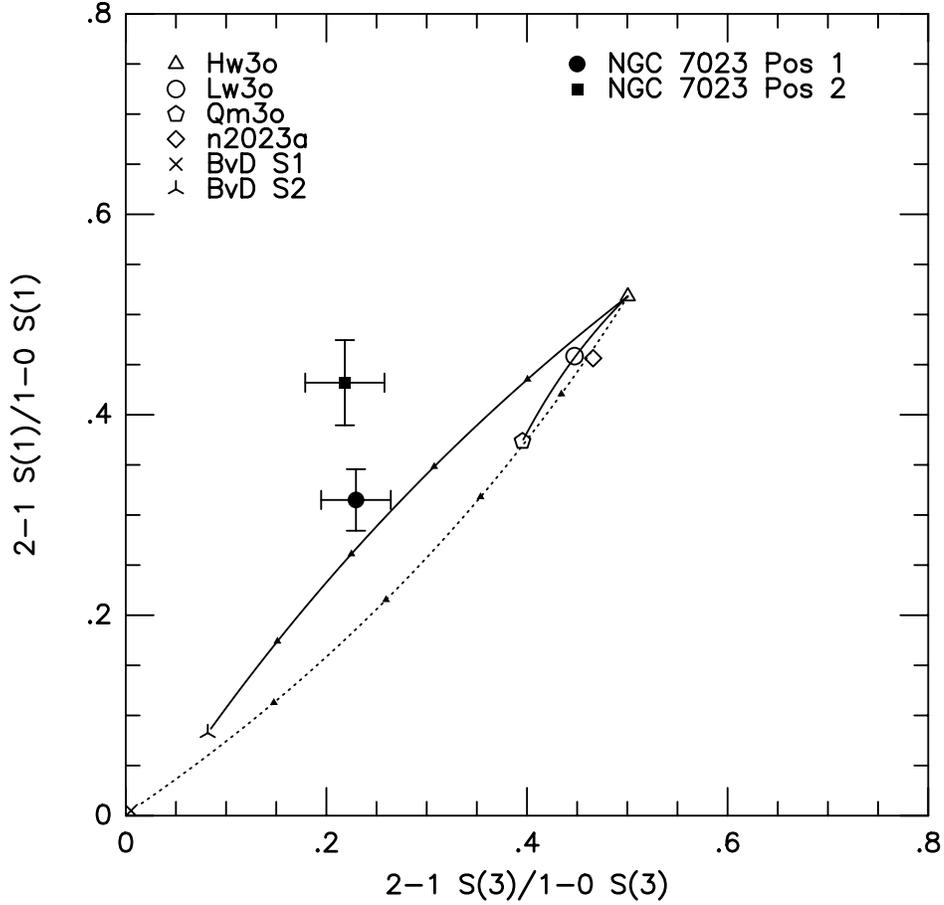}{5.0truein}{0}{70}{70}{-280}{0}
\caption{Ratio of the $2-1$ S(1)/$1-0$ S(1) lines vs. ratio of the
$2-1$ S(3)/$1-0$ S(3)
lines of \h2 for our observations of Position 1 and 2 in NGC 7023 (with
$1\sigma$ errorbars) and a selection of the
models listed in Table 4. All models from Table 4 not shown have similar
line ratios to the illustrated model at the same density but different UV
illumination and gas temperature. These ratios represent low, odd rotational
transitions where the effects of a thermalized component would be most
readily noticed
(see \S 6). Also shown are the way the line ratios change for model
Hw3o of Draine \& Bertoldi (1996) with the addition of the 1000 and 2000 K
thermal models of Black \& van Dishoeck (1987), and wth the high-density
model Qm3o (see \S 6). Tick marks denote 20, 40, 60, and 80\% contributions
of the thermal models. }
\end{figure}

\end{document}